\documentclass[superscriptaddress,amsmath,amssymb,aps,showkeys,showpacs,twoside,final,secnumarabic,nofootinbib]{revtex4-2}

\usepackage[paperwidth=205mm,paperheight=290mm,top=17mm,bottom=25mm,
inner=17mm,outer=17mm,
twoside]{geometry}

\usepackage{graphicx}
\usepackage[unicode=true,colorlinks=true,linkcolor=magenta, urlcolor=blue, citecolor = blue,breaklinks]{hyperref}

\begin{document}

\title{ Primordial black holes formation in inflationary $F(R)$ models\\ with scalar fields}

\def\addressa{Skobeltsyn Institute of Nuclear Physics,\\ Lomonosov Moscow State University,\\ Leninskie Gory 1, Moscow 119991,  Russia}

\author{\firstname{E.O.}~\surname{Pozdeeva}}
\email[E-mail: ]{pozdeeva@www-hep.sinp.msu.ru}
\affiliation{\addressa}
\author{\firstname{S.Yu.}~\surname{Vernov}}
\email[E-mail: ]{svernov@theory.sinp.msu.ru}
\affiliation{\addressa}

\begin{abstract}
We construct $F(R)$ gravity models with scalar fields to describe cosmological inflation and formation of primordial black holes (PBHs). By adding the induced gravity term and the fourth-order polynomial potential for the scalar field to the known $F(R)$ gravity model, and using a conformal transformation of the metric, we obtain a two-field chiral cosmological model. For some values of the model parameters, we get that the inflationary parameters of this model are in good agreement with the observations of the cosmic microwave background radiation obtained by the Atacama Cosmology Telescope. The estimation of PBH masses suggests that PBHs could be dark matter candidates.
\end{abstract}

\par
\keywords{Primordial black hole, inflation, modified gravity   \\[5pt]}

\maketitle

\section{Introduction}\label{intro}

A black hole is called primordial if it formed before the matter dominance epoch.
The hypothesis of the existence of primordial black holes (PBHs) is supported by an increasing amount of direct and indirect observations of black holes with masses beyond the astrophysical range, the occurrence of which is not explained by models of stellar collapse~\cite{Dolgov:2017aec,Dolgov:2020sov,DeLuca:2025fln}.
It is possible that a significant fraction, or even the entirety, of dark matter is not a new type of matter, but consists of PBHs~\cite{Khlopov:2008qy,Kamenshchik:2018sig,Green:2020jor,Carr:2020xqk,Carr:2020gox}.

The most popular PBH formation mechanism assumes the existence of the overdensities that are larger than a critical value forming during the accelerated expansion of the early Universe, known as cosmological inflation~\cite{Dolgov:1992pu,Ivanov:1994pa,Garcia-Bellido:1996mdl,Ketov:2021fww,Ozsoy:2023ryl}. These overdensities may form PBHs during the radiation dominated era.  Models that unify inflation with PBH formation require a violation of the slow-roll conditions during inflation~\cite{Germani:2017bcs,Ketov:2021fww,Ozsoy:2023ryl}. In single-field inflation models, PBH formation is associated with an ultra-slow-roll stage of inflation~\cite{Germani:2017bcs,Ezquiaga:2017fvi,Chataignier:2023ago,Kamenshchik:2024kay}.  It has been noted in Ref.~\cite{Kristiano:2022maq} that quantum loop corrections might invalidate some single-field models of inflation with PBH production.

There are many modified gravity models of cosmological inflation~\cite{Nojiri:2010wj,Capozziello:2011et,Odintsov:2023weg,Fazzari:2025nfr}. Models with nonminimally coupled scalar fields and $F(R)$ gravity models are classically equivalent to General Relativity (GR) with minimally coupled scalar fields~\cite{Maeda:1988ab}.
Using the Weyl transformation of the metric, one can transform  the original modified gravity description, known as the Jordan frame, into the GR description, known as the Einstein frame. In $F(R)$ inflationary models, the scalar field is identified with the inflaton having the clear gravitational origin as a physical excitation of the higher-derivative gravity (called scalaron). In $F(R)$ models and corresponding single-field inflationary models in the Einstein frame, PBH formation has been investigated in Refs.~\cite{Saburov:2023buy,Saburov:2024und}.

A $F(R)$ gravity model with a scalar field is equivalent to a two-field model in the Einstein frame. This two-field model has a non-standard kinetic term in the action, in other words, one gets a chiral cosmological model (CCM)~\cite{Chervon:1995jx,Starobinsky:2001xq,vandeBruck:2015xpa,Kaneda:2015jma,Karananas:2016kyt,Pi:2017gih,Chervon:2019nwq,Fomin:2020caa,Ivanov:2021ily,Geller:2022nkr,Pozdeeva:2024lah}. Two-field CCMs are actively used to describe inflation suitable for the PBH formation~\cite{Pi:2017gih,Braglia:2020eai,Gundhi:2020zvb,Gundhi:2020kzm,Braglia:2020eai,Ketov:2021fww,Braglia:2022phb,Geller:2022nkr,Cheong:2022gfc,Kawai:2022emp,Pozdeeva:2024hfq,Wang:2024vfv,Kim:2025dyi,Wang:2025dbj}.

The most known example of a $F(R)$ gravity model with a scalar field is the Higgs-$R^2$ inflationary model, which includes a quadratic curvature term and a nonminimal coupling between the Higgs boson and gravity~\cite{Ema:2017rqn,Wang:2017fuy,He:2018gyf,Gorbunov:2018llf,Enckell:2018uic,Bezrukov:2019ylq,Ema:2019fdd,He:2020ivk,Ema:2020evi,Cado:2024von,Kim:2025ikw}. This model has been used to investigate the formation of PBHs in Ref.~\cite{Gundhi:2020zvb}. In many two-field models, one scalar field plays a role of inflaton in the beginning of inflation and another field plays the same role at the end. The investigations of such inflationary models with two stages of inflation show that energy density perturbations at the time corresponding to the transition between two inflationary stages can be so large that lead to PBH production~\cite{Garcia-Bellido:1996mdl,Pi:2017gih,Gundhi:2020kzm,Braglia:2020eai,Ketov:2021fww,Braglia:2022phb,Cheong:2022gfc,Kawai:2022emp}.

The first and most well-known $F(R)$ gravity inflationary model is the Starobinsky model, proposed in 1980~\cite{Starobinsky:1980te} (see also Refs.~\cite{Mukhanov:1981xt,Vilenkin:1985md,Mijic:1986iv,Maeda:1987xf,Ketov:2025nkr,Linde:2025pvj}). After the Starobinsky model, many $F(R)$ gravity models of inflation have been proposed~\cite{Berkin:1990nu,Saidov:2010wx,Kaneda:2010ut,Kaneda:2010qv,Ketov:2010qz,Huang:2013hsb,Motohashi:2014tra,Odintsov:2016vzz,Miranda:2017juz,Waeming:2020rir,Rodrigues-da-Silva:2021jab,Ivanov:2021chn,Pozdeeva:2022lcj,Oikonomou:2025qub,Fazzari:2025nfr}. New observation data obtained by the Atacama Cosmology Telescope (ACT)~\cite{ACT:2025fju,ACT:2025tim} combined with DESI 2024 results~\cite{DESI:2024mwx} have merely $\sim 2\sigma$ tension with the predictions of the Starobinsky model~\cite{Kallosh:2025ijd}. So, it is reasonable to construct and investigate $F(R)$ inflationary models fitting the most recent observation data~\cite{Addazi:2025qra,Gialamas:2025ofz,Ketov:2025cqg,Ivanov:2025nsx,Odintsov:2025jky,Odintsov:2025eiv}.

In this paper, we propose a $F(R,\chi)$ inflationary model with the scalar field $\chi$. We compare $F(R)$ gravity inflationary models that have been constructed or developed to fit the ACT data and show that the model proposed in Ref.~\cite{Ivanov:2025nsx} is the most suitable for our proposals. We add the scalar field with the induced gravity term and the fourth-order polynomial potential to this $F(R)$ model. Using conformal transformation of the metric, we get a CCM model with two scalar fields. We analyze the behaviour of scalar fields during inflation by numerical calculations for different values of the model parameters and demonstrate that the constructed inflationary model does not contradict to the recent ACT/DESI observation data and is suitable for PBH formation. The estimation of PBH masses shows that PBHs could be dark matter candidates.

\section{$F(R,\chi)$ gravity models and two-field models}

We consider a generic $F({R})$ gravity model with a scalar field~$\chi$, described by
\begin{equation}
S_{R}=\int d^4x\sqrt{-{g}}\left[F({R},\chi)-\frac12{g}^{\mu\nu}\partial_{\mu}\chi\partial_{\nu}\chi \right],
\label{actionFR}
\end{equation}
where $F({R},\chi)$ is a nonlinear double differentiable function.
Action (\ref{actionFR}) can be rewritten in the following form:
\begin{equation}
\label{SJFR}
S_{J}=\int d^4x\sqrt{-{g}}\left[F_{,\sigma}{R}-V-\frac12{g}^{\mu\nu}\partial_{\mu}\chi\partial_{\nu}\chi \right],
\end{equation}
where $F_{,\sigma}= \frac{\partial F}{\partial \sigma}$, $V\equiv F_{,\sigma}\sigma - F$. Varying action (\ref{SJFR}) with respect to $\sigma$, it is straightforward to get the equation $F_{,\sigma\sigma}(\sigma-{R})=0$ and to recover the original action (\ref{actionFR}).

For metric gravity models, the conformal transformation of the metric:
\begin{equation}
\label{gJgE}
\tilde{g}_{\mu\nu}=\frac{2F_{,\sigma}}{M^2_\mathrm{Pl}}{g}_{\mu\nu},
\end{equation}
gives  the following CCM models in the Einstein frame:
\begin{equation}
\label{FRSE}
S_{E}=\int d^4x\sqrt{-\tilde{g}}\left[\frac{M^2_\mathrm{Pl}}{2}\tilde{R}-\frac{3M^2_\mathrm{Pl}F_{,\sigma\sigma}^2}{4F_{,\sigma}^2}{\tilde{g}^{\mu\nu}}\partial_\mu\sigma\partial_\nu\sigma
-\frac{M^2_\mathrm{Pl}}{4F_{,\sigma}}{\tilde{g}^{\mu\nu}}\partial_\mu\chi\partial_\nu\chi-V_E\right],
\end{equation}
where
\begin{equation*}
V_E= \frac{M^4_\mathrm{Pl}}{4F_{,\sigma}^2}\left(F_{,\sigma}\sigma-F\right)\,.
\end{equation*}

Introducing
\begin{equation}
\label{phisigmaFr}
    \phi=\sqrt{\frac{3}{2}}M_\mathrm{Pl}\ln\left(\frac{2F_{,\sigma}}{M_\mathrm{Pl}^2}\right)\,,
\end{equation}
we obtain
 \begin{equation}
\label{SE2}
S_{E}=\int d^4x\sqrt{-\tilde{g}}\left[\frac{M^2_\mathrm{Pl}}{2}\tilde{R}-\frac{1}{2}\tilde{g}^{\mu\nu}\partial_{\mu}\phi\partial_{\nu}\phi-\frac{y}{2}\tilde{g}^{\mu\nu}\partial_{\mu}\chi\partial_{\nu}\chi-V_E(\phi,\chi)\right],
\end{equation}
where
\begin{equation}
\label{yphi}
y= \frac{M^2_\mathrm{Pl}}{2F_{,\sigma}}=\mathrm{e}^{-\sqrt{\frac{2}{3}}\,\frac{\phi}{M_\mathrm{Pl}}},\qquad V_E(\phi,\chi)=y^2(\phi)V(\sigma(\phi,\chi),\chi).
\end{equation}

\section{Evolution equations and inflation}

\subsection{Exact evolution equations}

In the spatially flat Friedmann--Lema\^{i}tre--Robert\-son--Walker metric with the interval
\begin{equation*}
{ds}^{2}={}-{dt}^2+a^2(t)\left(dx_1^2+dx_2^2+dx_3^2\right),
\end{equation*}
the model (\ref{SE2}) has the following evolution equations~\cite{Chervon:2019nwq,Pozdeeva:2024hfq}:
\begin{equation}
    \label{H2}
H^2=\frac{1}{6M_\mathrm{Pl}^2}\left(X^2+2V_E\right)\,,
\end{equation}
\begin{equation}
    \label{dH}
\dot H={}-\frac{X^2}{2M_\mathrm{Pl}^2} \,,
\end{equation}
where dots denote the time derivatives, $X\equiv\sqrt{{\dot{\phi}}^2+y\,{\dot{\chi}}^2}$ and the Hubble parameter $H$ is the logarithmic derivative of the scale factor: $ H=\dot a/{a}$.

The field equations are
\begin{equation}
\label{equphi}
\ddot{\phi}+3H\dot{\phi}-\frac12\, \frac{dy}{d\phi}\,\dot{\chi}^2+\frac{\partial V_E}{\partial \phi}=0\,,
\end{equation}
\begin{equation}
\label{equxi}
\ddot{\chi}+3H\dot{\chi}+\frac{1}{y}\, \frac{dy}{d\phi}\,\dot{\chi}\dot{\phi}+\frac{1}{y}\,\frac{\partial V_E}{\partial \chi}=0\,.
\end{equation}

It is suitable to consider the e-folding number
$N=\ln(a/a_{i})$, where $a_{i}$ is a constant, as an independent variable during inflation. We choose a such value of $a_i$ that the inflationary parameters are calculated at $N=0$.

Using the relation $\frac{d}{dt}=H\,\frac{d}{dN}$, Eqs.~\eqref{H2}
and \eqref{dH} can be rewritten as follows
\begin{equation}
    \label{a2N} H^2=\frac{2V_E}{6M_\mathrm{Pl}^2-{\phi^\prime}^2-y{\chi^\prime}^2}\,,
\end{equation}
\begin{equation}\label{dHdN}
 H'={}-\frac{H}{2M_\mathrm{Pl}^2}\left[{\phi^\prime}^2+y{\chi^\prime}^2\right]\,,
\end{equation}
where primes denote derivatives with respect to $N$.

The standard slow-roll parameters in the Einstein frame are defined as~\cite{Liddle:1994dx}:
\begin{eqnarray}
  \varepsilon &=&{}-\frac{\dot{H}}{H^2}={}-\frac{H^\prime}{H}=\frac{1}{2M_\mathrm{Pl}^2}\left[{\phi^\prime}^2+y{\chi^\prime}^2\right], \\
  \eta\label{eta}
  &=&{}-\frac{\ddot{H}}{2H\dot{H}}={}-\frac{1}{2}\frac{\left(H^2\right)^{\prime\prime}}{\left(H^2\right)^{\prime}}=\varepsilon-\frac{\varepsilon^\prime}{2\varepsilon}~.\label{eta}
\end{eqnarray}

Using Eqs.~(\ref{a2N}) and (\ref{dHdN}), we eliminate $H^2$ and $H'$ from the field equations and obtain the following system of two second-order differential equations:
\begin{equation}
\label{DYNSYSTEMN}
\begin{split}
    \phi''=&(\varepsilon-3)\phi'+\frac12\,\frac{dy}{d\phi}\,{\chi'}^2-\frac{6M_\mathrm{Pl}^{2}-y{\chi'}^2-{\phi'}^2}{2}\,\frac{\partial \ln(V_E)}{\partial \phi}\,,\\
    {\chi}''=&(\varepsilon-3)\chi'+\frac{2}{\sqrt{6\,}M_\mathrm{Pl}}\chi'\phi'-\frac{6M_\mathrm{Pl}^2-{\phi'}^2-y{\chi'}^2}{2y}\frac{\partial \ln(V_E)}{\partial \chi}\,.\\
\end{split}
\end{equation}

\subsection{Slow-roll and ultra-slow-roll regimes of inflation}

Using Eq.~(\ref{DYNSYSTEMN}), we obtain
\begin{equation}
\varepsilon'=2\varepsilon(\varepsilon-3)-\frac{1}{M^2_\mathrm{Pl}H^2}V_E'\,.
\end{equation}

So, Eq.~(\ref{eta}) gives
\begin{equation}\label{eta2}
    \eta=3+\frac12\left(\frac{3}{\epsilon}-1\right)\frac{d\ln
V_E}{dN}=3 + \frac{V_E'}{2\varepsilon M^2_\mathrm{Pl}H^2}=3-\frac{V_E'}{M^2_\mathrm{Pl}\left(H^2\right)'}\,.
\end{equation}

Absolute values of both slow-roll parameters, $\epsilon$ and $\eta$, should be smaller than one in the slow-roll regime.
When $\epsilon$ becomes equal to one, inflation as an accelerated expansion of the Universe stops, therefore, only $|\eta|$ can be greater than one during inflation.
The ultra-slow-roll regime occurs when $\eta\approx 3$.

Our goal is to propose a model that realizes so-called two-stage inflationary scenario~\cite{Ketov:2021fww,Braglia:2022phb,Pozdeeva:2024hfq,Wang:2024vfv}. In the first stage, the scalar field $\chi$ remains almost constant and only the field $\phi$ evolves. This stage satisfies the slow-roll conditions, and the inflationary parameters can be calculated using standard formulae for the slow-roll approximation. The second stage of inflation corresponds to the evolution of the scalar field $\chi$. The slow-roll regime is violated when the first stage of the inflation ends. During a few e-foldings the absolute value of the parameter $\eta$ can be greater than one. After this, the slow-roll approximation recovers. This violation of the slow-roll approximation is a necessary condition for PBH formation.

 At the ultra-slow-roll regime, we have a nearly-inflection point, at which $V_{E}'\approx 0$ (see Ref.~\cite{Braglia:2022phb,Pi:2017gih}). To describe this point~\cite{Kamenshchik:2024kay} we use the slow-roll parameter $\eta$. If $V_E'\approx 0$, then Eq.~(\ref{eta2}) gives $\eta\approx3$. We use the supposition that the transition from the first stage of inflation  to
the second stage leads to grow of energy density perturbations leading to PBH formation at the
movement when perturbations with wavenumber around $k_{*}$ re-enter the horizon $k_{*}=a_{*}H_{*}=a_{re}H_{re}=k_{re}$ \cite{Garcia-Bellido:1996mdl,Pi:2017gih}. Modes of perturbations can re-enter the horizon in different stage of the universe
evolution. We works in the supposition that it is taking place
during radiation dominant stage and e-folding numbers at which  PBH
formation is possible is very close to the end of second stage of
inflation~\cite{Pi:2017gih}.

The mass of PBHs depends on the duration of the second stage, $N_{e}-N_{*}$, and the
value of the Hubble parameter at the end of inflation $H_{e}$.
To estimate the mass of PBHs
 we apply the formula from Refs.~\cite{Pi:2017gih,Frolovsky:2022qpg} in  the
form obtained in Ref.~\cite{Pozdeeva:2024hfq}:
\begin{equation}
M_{PBH}\simeq\frac{M_\mathrm{Pl}^2}{H_{e}}\exp\left(2(N_{e}-N_{*})\right),
\end{equation}
where $N_{e}$ is the total duration of inflation, $N_*$ is the minimal value of $N$, at which $\eta(N_{*})=3$.

\section{$F(R,\chi)$ inflationary model suitable for PBH formation}

\subsection{$F(R)$ inflationary models}

The Starobinsky inflationary model is described by the following action,
\begin{equation}
\label{starS}
    S_{\rm Star.} = \frac{M^2_\mathrm{Pl}}{2}\int d^4 x \sqrt{-g } \left( {R} + \frac{1}{6m^2}{R}^2\right)~,
\end{equation}
with only one parameter $m\sim 10^{-5}M_\mathrm{Pl}$, which is the inflaton mass.  The inflationary  parameters $n_s$ and $r$ do not depend on $m$, but depend on the number of e-foldings during inflation $N_i$:
\begin{equation}
n_s=1-\frac{2}{N_i}+{\cal{O}}(N_i^{-2}),\qquad r=\frac{12}{N_i^2}+{\cal{O}}(N_i^{-3})~.
\end{equation}
In particular, $n_s=0.964$ corresponds to $N_i\approx 55$, whereas $n_s=0.974$ corresponds to $N_i\approx 77$.

This means that the Starobinsky inflation nicely fits the CMB observations by the Planck/BICEP collaborations~\cite{Planck:2018jri,BICEP:2021xfz,Galloni:2022mok},
\begin{equation} \label{PLB}
n_s = 0.9651 \pm 0.0044\,,
 \end{equation}
but deviates from the latest ACT/DESI data~\cite{ACT:2025fju,ACT:2025tim,DESI:2024mwx},
\begin{equation}\label{ACT}
n_s = 0.9743 \pm 0.0034\,.
\end{equation}

Note that the ACT/DESI data does not significantly change the upper bound on the tensor-to-scalar ratio $r$ and the value of the amplitude of scalar perturbations $A_s$,
\begin{equation}
\label{Inflparamobserv}
A_s=(2.10\pm 0.03)\times 10^{-9}\qquad {\rm and} \qquad  r < 0.028~.
\end{equation}

To construct a $F(R,\phi)$ inflationary model suitable for primordial black hole formation, we first need to find an $F(R)$ inflation model that fits the ACT/DESI data.

The Starobinsky model has several important properties that must be satisfied when one constructs a well-behaved $F(R)$ inflationary model. First of all, one needs to avoid graviton as a ghost and scalaron (inflaton) as a tachyon. These conditions put the following restrictions on $F({R})$ function~\cite{Starobinsky:2007hu}:
\begin{equation}
\label{restr2}
F_{,{R}}\equiv\frac{dF}{d{R}} >0 \quad {\rm and} \quad \frac{d^2F}{d{R}^2} >0\,.
\end{equation}
In the Starobinsky model, these restrictions are satisfies for all ${R}>-3m^2$.

It has been shown in Refs.~\cite{Skugoreva:2014gka,Vernov:2021hxo} that de Sitter solutions correspond to maxima and minima of the effective potential
\begin{equation}
\label{Ve}
V_{eff}(\sigma)= \frac{M^4_\mathrm{Pl}\left(F_{,\sigma} \sigma-F\right)}{4F_{,\sigma}^2}\,,
\end{equation}
where $\sigma$ is a scalar field associated with ${R}$ as in action~(\ref{SJFR}).  In the Starobinsky model, $V_{eff}(\sigma)$ is a monotonically increasing function for all $\sigma\geqslant 0$. It means that we can use any sufficiently large value of ${R}$ as an initial condition for the inflationary trajectory, therefore, there is no problem with fine-tuning of the initial data~\cite{Mishra:2019ymr}.

A few $F(R)$ gravity models have been proposed or developed~\cite{Gialamas:2025ofz,Addazi:2025qra,Ketov:2025cqg,Ivanov:2025nsx,Odintsov:2025jky,Odintsov:2025eiv} to fit the ACT/DESI data.
The simplest way to modify the Starobinsky model is to add $c_n {R}^n$ terms, where $c_n$ are constants and natural number $n>2$ (see Refs.~\cite{Berkin:1990nu,Saidov:2010wx,Kaneda:2010qv,Rodrigues-da-Silva:2021jab,Ivanov:2021chn,Addazi:2025qra,Gialamas:2025ofz,Ketov:2025cqg}).
The $c_n R^n$ term dominates at large $R$, so, one needs some fine-tuning of initial data because for large ${R}$ either $F_{,{R}}<0$ (at $c_n<0$) or an unstable de Sitter solution exists (at $c_n>0$). Note that the modifications of the Higgs$-R^2$ inflation by $R^3$ term~\cite{Pi:2017gih,Kim:2025dyi,Modak:2025bjv} have the same problem. Model with ${R}^{3/2}$ correction proposed in Ref.~\cite{Ivanov:2021chn} can describe only minimally possible value of $n_s$ (see Ref.~\cite{Addazi:2025qra} for detail). Also  in this model, the flat space-time with ${R}=0$ corresponds to singularity in $F_{,{R}}$ function. The same problem appears in models~\cite{Motohashi:2014tra,Waeming:2020rir,Odintsov:2025jky,Odintsov:2025eiv} with $F({R})\sim{R}+{R}^p$, where $p<2$ is a real number. To get a monotonically increasing effective potential at a finite $F_{,{R}}(0)>0$ the model with $(R+R_0)^{3/2}$ term, where a constant $R_0>0$, has been proposed in Ref.~\cite{Pozdeeva:2022lcj}. This model  as well as model with ${R}^{3/2}$ term can describe only minimally possible value of $n_s$.

Only the $F(R)$ model proposed in Ref.~\cite{Ivanov:2025nsx} has all above-mentioned important properties and is in agreement with  the ACT/DESI data. We add a scalar field to this model and consider a possibility of the PBH production in the obtained two-field model.

\subsection{Two-field CCM}

We consider the following modified gravity model:
\begin{equation}\label{ModelF}
F({R},\chi)=\frac{M^2_\mathrm{Pl}}{2}\, \left[\left( 1+X(\chi) \right)  \left(
1-\frac{1}{3\,{\delta}}\right) R+\frac{1}{3\delta}
\left( R+{\frac {{m}^{2}}{ \delta}} \right) \ln \left( 1+{\frac
{\delta\,R}{{m}^{2}}}\right)-U(\chi)m^2\right]\,,
\end{equation}
where $\delta$ is a dimensionless positive constant, $X(\chi)$ and $U(\chi)$ are dimensionless differentiable functions of the scalar field $\chi$.

We choose the following fourth-order polynomial function~$U(\chi)$ and such function $X(\chi)$ that $X(\chi)R$ is the induced gravity term,
\begin{equation}
X(\chi)=c\frac{\chi^2}{\chi_0^2},\qquad U(\chi)=U_0\left[\left(1-\frac{\chi^2}{\chi_0^2}\right)^2-d\frac{\chi}{\chi_0}\right],
\end{equation}
where $c$, $d$, $U_0$, and $\chi_0>0$ are constants.
The original $F(R)$ model~\cite{Ivanov:2025nsx} corresponds to $X(\chi)\equiv 0$ and $U(\chi)\equiv 0$.

For $\chi=0$ and small $R$, we get the Starobinsky inflationary model with a cosmological constant at any value of the parameter $\delta$,
\begin{equation}
F|_{\chi=0}=\frac{M^2_\mathrm{Pl}}{2}\left(-U_0m^2+R+\frac{R^2}{6m^2}+{\cal{O}}\left(R^3\right)\right).
\end{equation}

A nice feature of the model (\ref{ModelF}) is the existence of the potential $V_E(\phi,\chi)$ in the analytic form.
Using Eq.~\eqref{phisigmaFr}, we get the
following relation:
\begin{equation}
\sigma=\frac{{m}^{2}}{\delta} \left( {\exp\left(-{\frac{(3\,\delta-1) c{\chi}^{2}y+3\,
\delta\chi_0^{2}(y-1)}{\chi_0^{2}y}}\right)}-1
\right)\,.
\end{equation}
So, the potential in the Einstein frame can be presented as
\begin{equation}
\begin{split}
V_E&=\frac{M^2_\mathrm{Pl}{m}^{2}\,y}{2{\delta}^{2}}\left[\frac{y}{3}\,\exp\left(-{\frac{(3\,\delta-1) c{\chi}^{2}y+3\,
\delta\chi_0^{2}(y-1)}{\chi_0^{2}y}}\right)+U_0\,{\delta}^{2}y\frac{\chi^{4}}{\chi_0^{4}}\right.\\
&\left.{}+y\frac{{\chi}^{2}}{\chi_0^{2}}\left(c\delta-\frac{c}{3}-2{U_0}\,{\delta}^{2}\right)-{U_0}\,d{\delta}^{2}y\frac{{\chi}}{\chi_0}+\left({U_0}\,{\delta}^{2}-\frac{1}{3}+\delta
\right)y-\delta
\right]\,,
\end{split}
\end{equation}
where $y$ is defined by Eq.~(\ref{yphi}).

\subsection{Inflation at different values of the model parameters}

To get suitable inflationary scenarios we solve numerically system (\ref{DYNSYSTEMN}) for different values of the model parameters.
A notable feature of system (\ref{DYNSYSTEMN}) is that the potential $V_E$ appears only as the first derivative of its logarithm. It means that solutions of this system do not depend on the parameter $m$. So, the scalar spectral index $n_s$ and the tensor-to-scalar ratio $r$ do not depend on $m$. This parameter is defined by the observation value of the amplitude of scalar perturbations $A_s$.

At the first stage of inflation $\chi\approx 0$ and only $\phi$ changes. This stage is in the slow-roll regime, because both slow-roll parameters are smaller than one.
It allows us to fix $\chi=0$ and to consider this stage as a single-field slow-roll inflationary trajectory. In particular, we use the standard slow-roll formulae to connect the inflationary and slow-roll parameters~\cite{Liddle:1994dx}:
\begin{equation}\label{nsr}
    n_s=1-6\epsilon+2\eta,\qquad r=16\epsilon,\qquad A_s=\frac{2H^2}{\pi^2M_\mathrm{Pl}^2r}\,.
\end{equation}

For the corresponding $F(R)$ model, the following estimations for the inflationary parameters as functions of $N$ have been found~\cite{Ivanov:2025nsx}:
\begin{equation}
\label{ns_approx}
n_s \approx 1 - \frac{8\sqrt{2 \delta} N}{3\tan \left(\frac43\sqrt{2 \delta}  N\right)} \approx 1 - \frac{2}{N}+\frac{64\delta}{27} N+\dots,
\end{equation}
\begin{equation}
\label{r_approx}
    r \approx \frac{64 \delta}{3\sin^2\left(\frac43\sqrt{2 \delta}  N\right)}\approx \frac{12}{N^2}+\frac{128\delta}{9}+\frac{4096}{405}\delta^2 N^2+\dots\,.
\end{equation}

For this model and $50<N<60$, suitable values of $\delta$ belong the interval $2.7\times 10^{-5}<\delta< 1.2\times 10^{-4}$~\cite{Ivanov:2025nsx}. In our model, we choose $N$ equal to the number of e-folding during only the first stage of inflation, so $35< N < 40$. Using Eq.~(\ref{ns_approx}), we obtain $3.1\times 10^{-4}<\delta< 3.7\times 10^{-4}$ for $N=35$ and $2.1\times 10^{-4}<\delta< 2.6\times 10^{-4}$ for $N=40$. Thus, the suitable interval for the parameter $\delta$ is $2.1 \times 10^{-4}<\delta< 3.7 \times 10^{-4}$. It should be noted that this estimation has been obtained for $U_0 = 0$, but it can be used with reasonable accuracy for suitable non-zero values of the parameter $U_0$.

\begin{figure}
\includegraphics[scale=0.23]{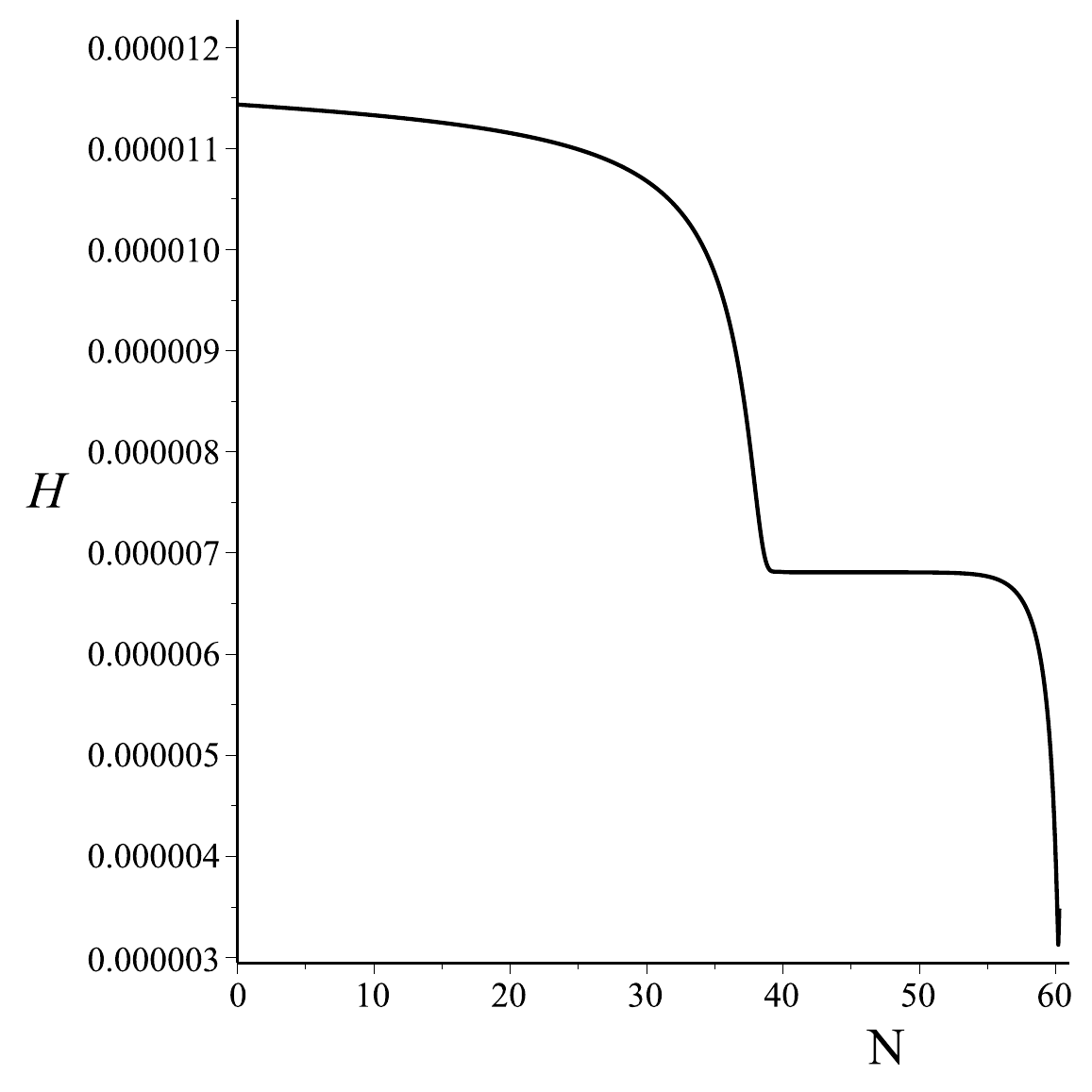} \
\includegraphics[scale=0.23]{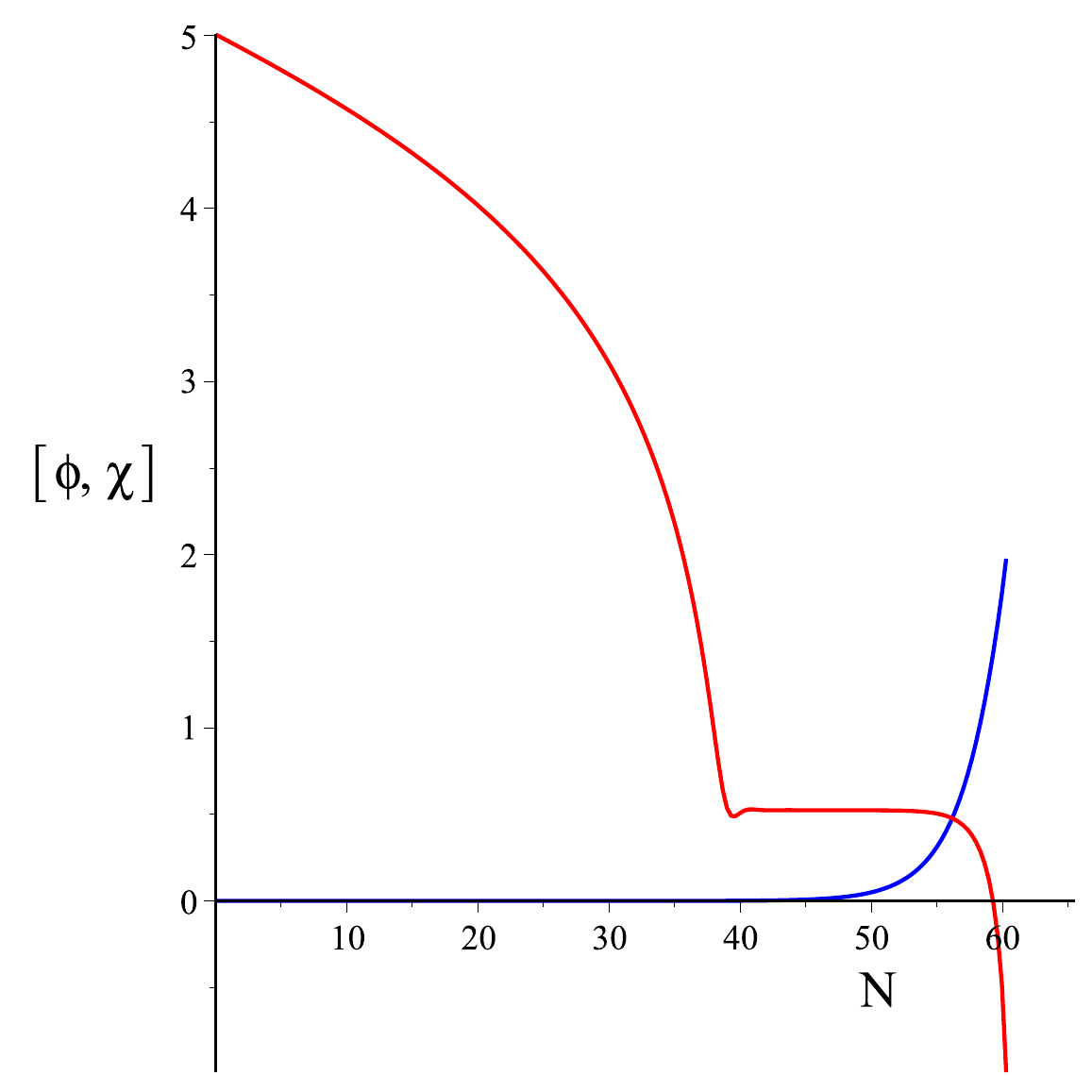}
\includegraphics[scale=0.29]{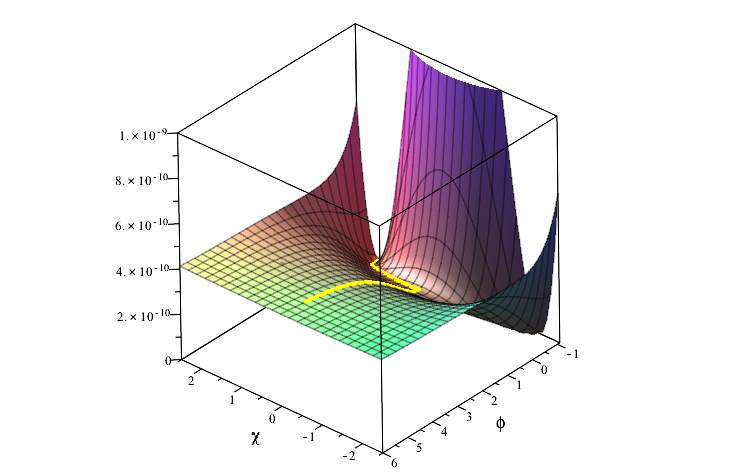}
\caption{The Hubble function $H(N)$ (left), the fields $\phi(N)$ [a red curve] and $\chi(N)$ [a blue curve] (center), and the potential $V_E$ with a trajectory (left). The model parameters are given in (\ref{modelparam}).}
\label{Trajectory}
\end{figure}

\begin{figure}
\includegraphics[scale=0.27]{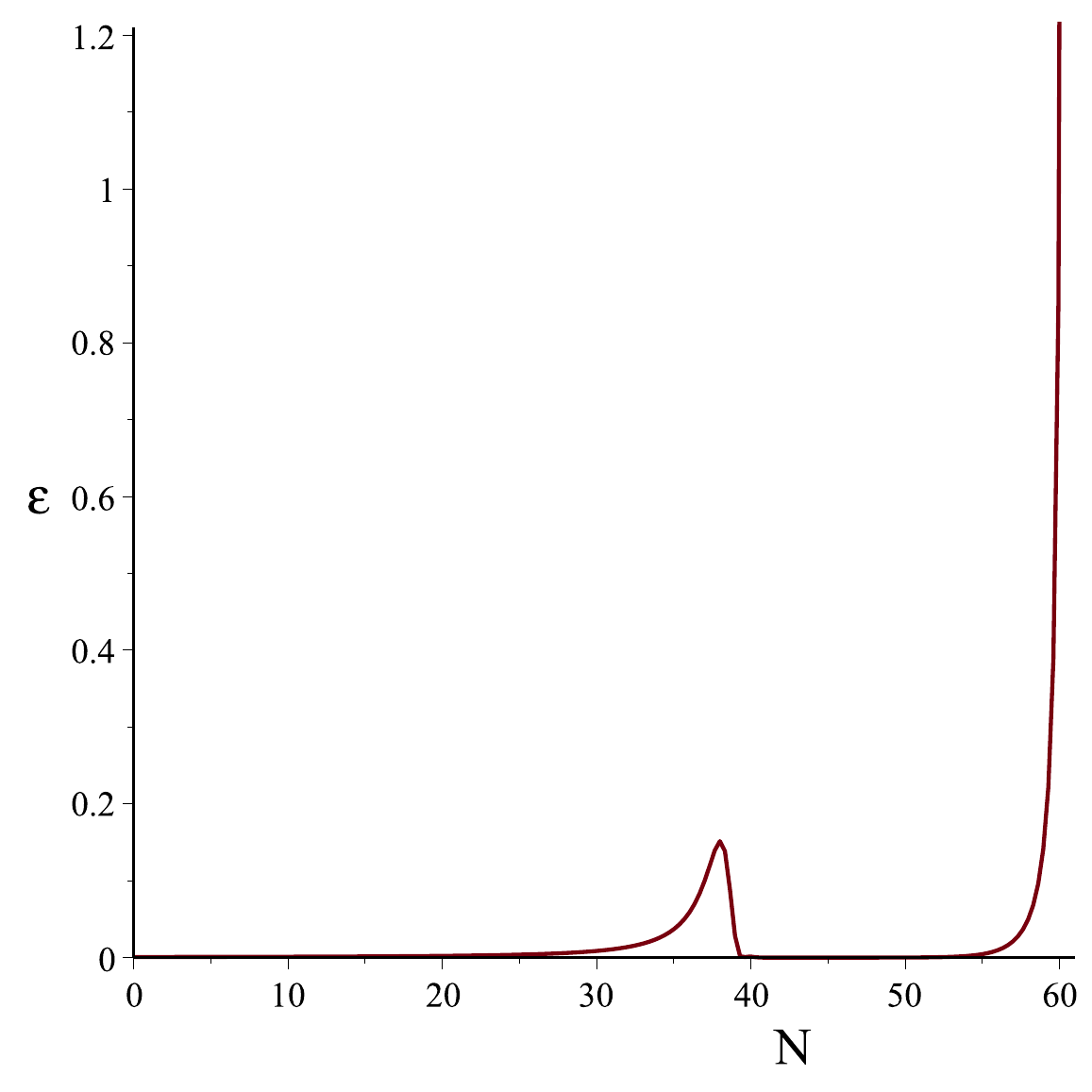} \
\includegraphics[scale=0.27]{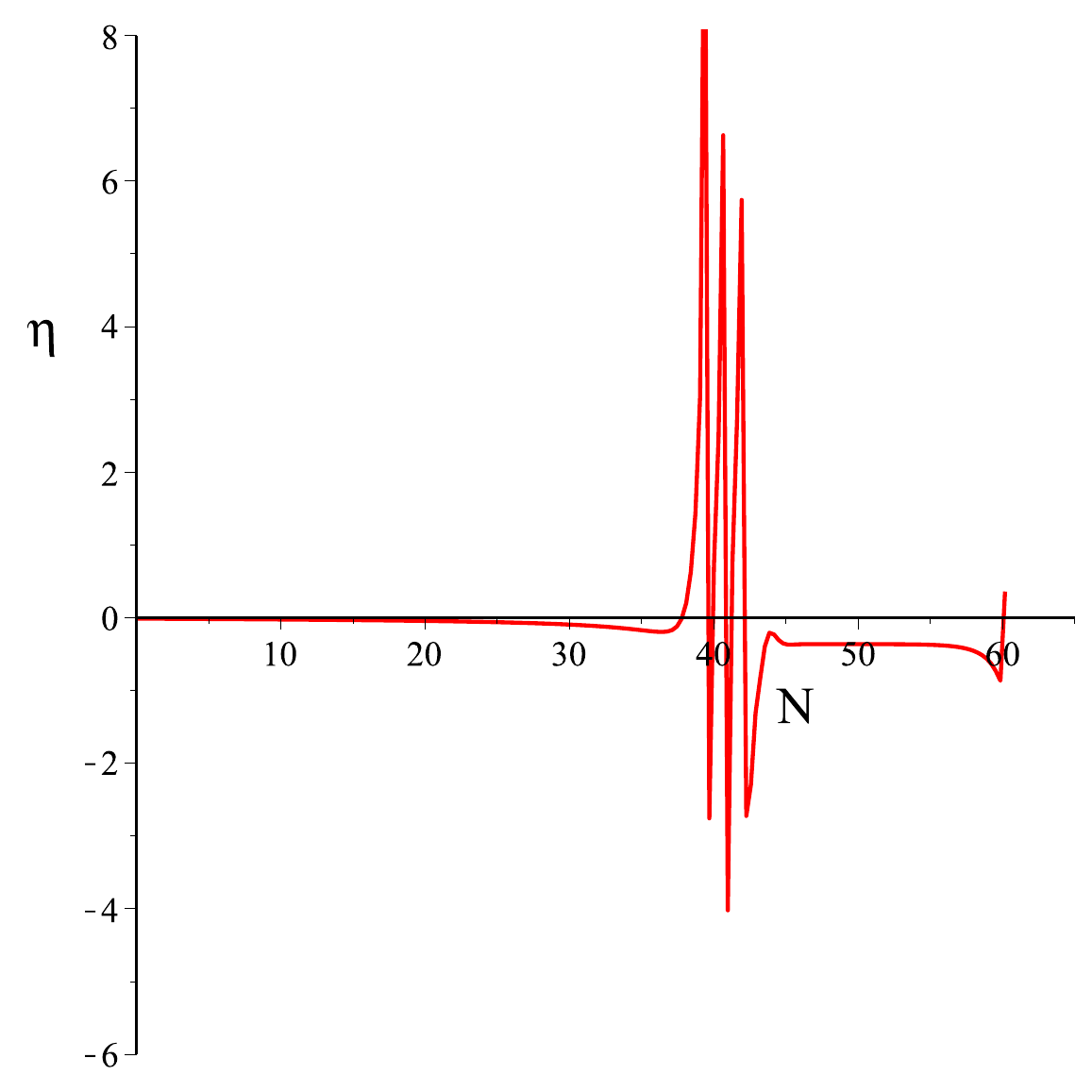} \
\includegraphics[scale=0.27]{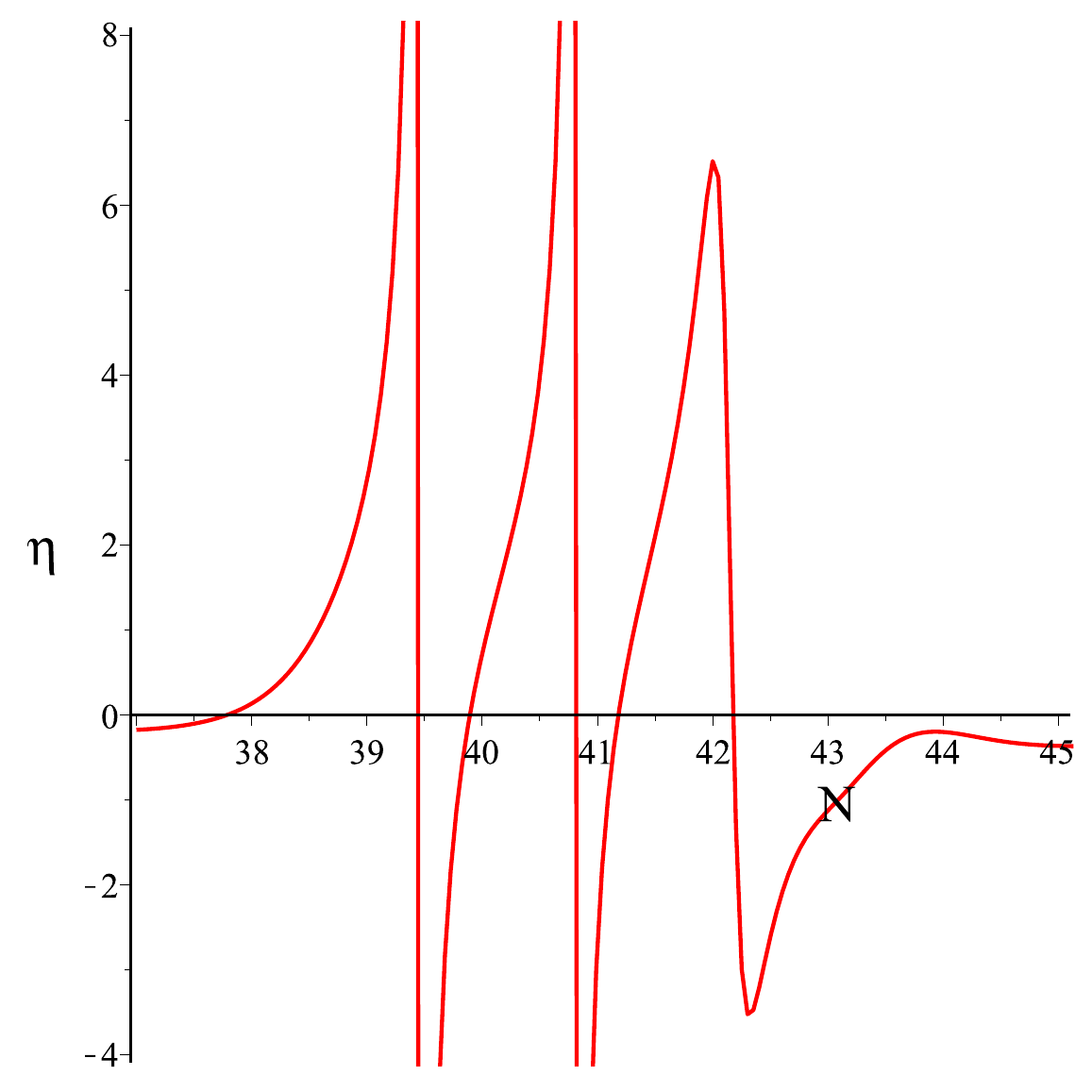}
\caption{The evolution of the slow-roll parameters $\epsilon(N)$ (left) and $\eta(N)$ (center and right) during inflation. The model parameters are given in (\ref{modelparam}).}
\label{slrparamN}
\end{figure}

\begin{figure}
\includegraphics[scale=0.27]{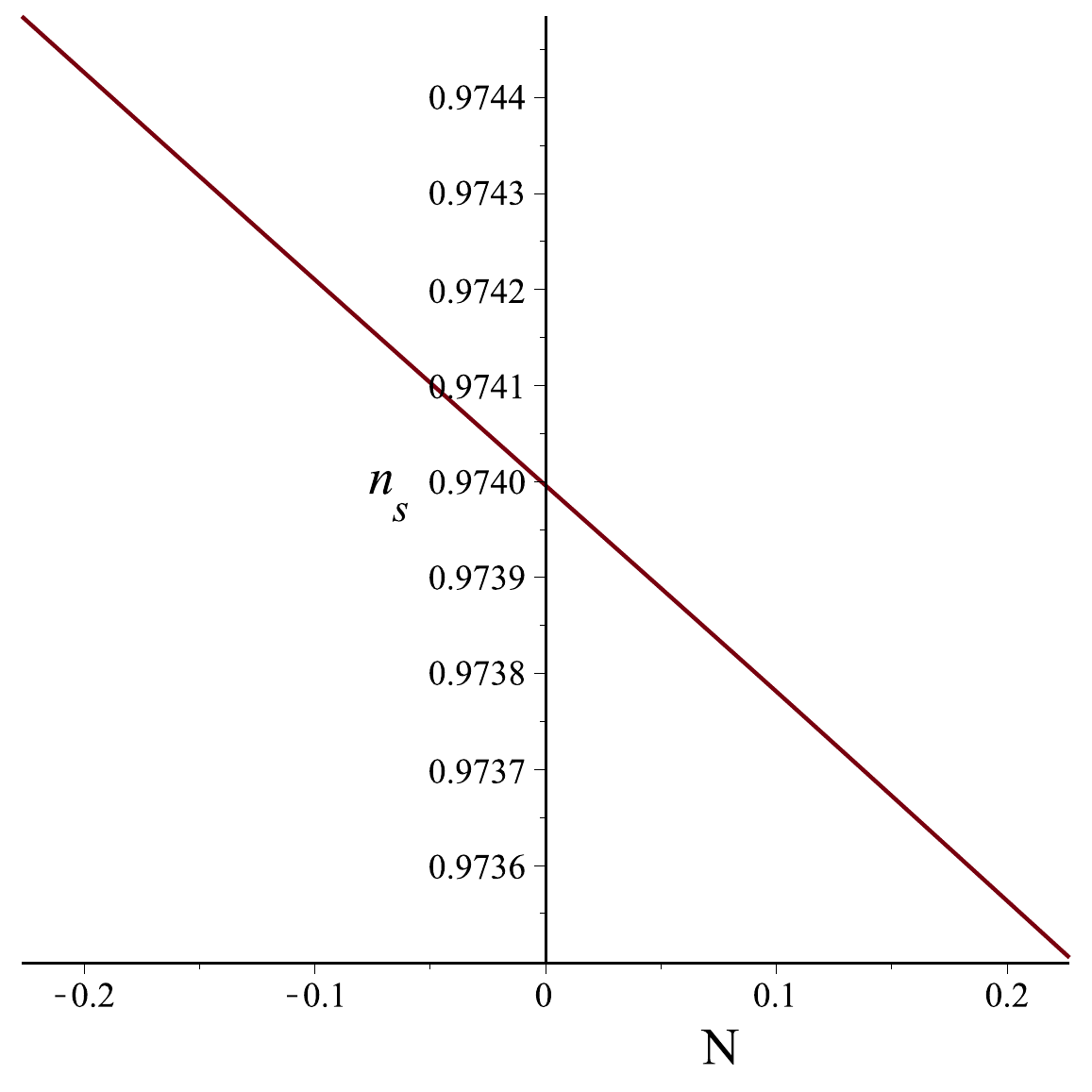} \
\includegraphics[scale=0.27]{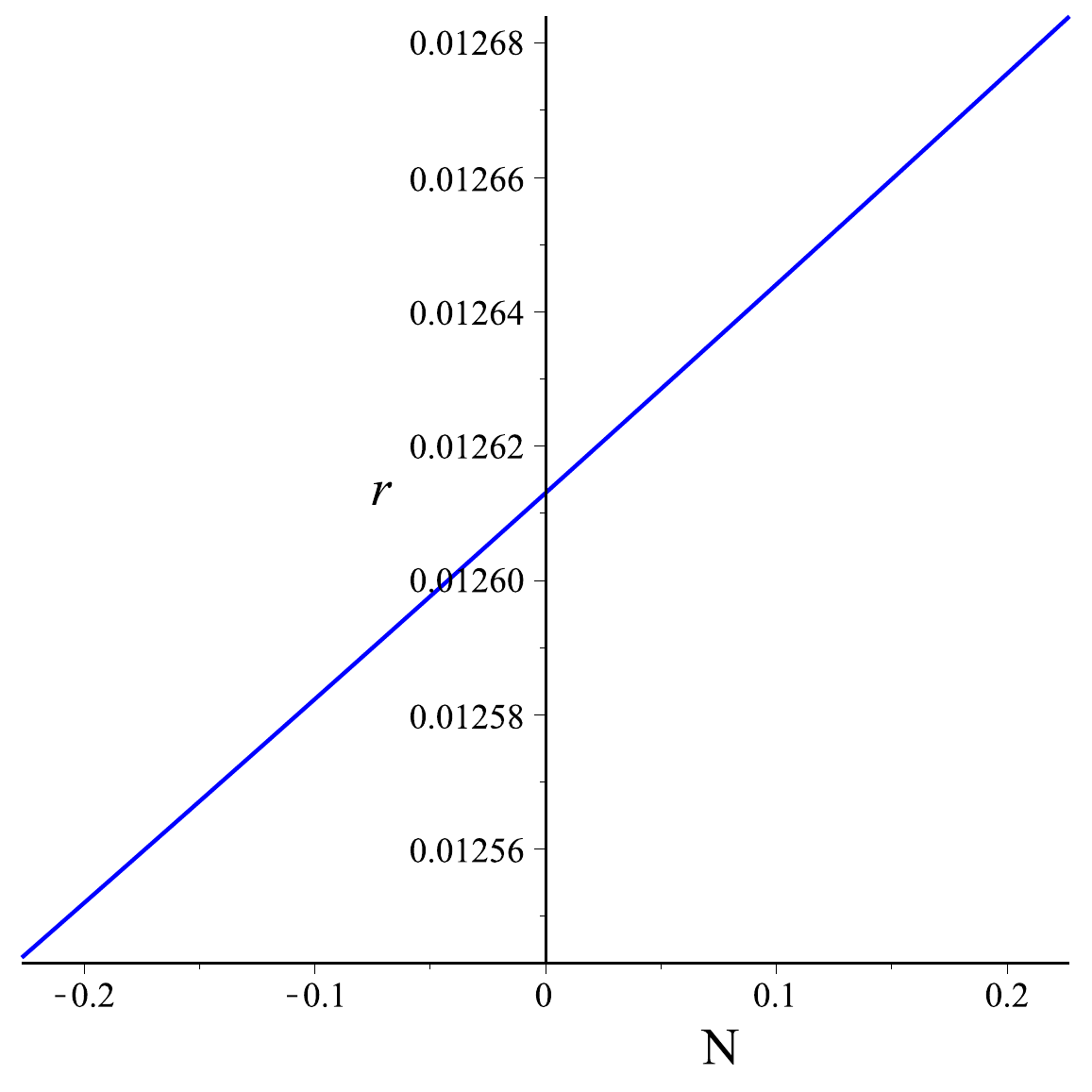} \
 \includegraphics[scale=0.27]{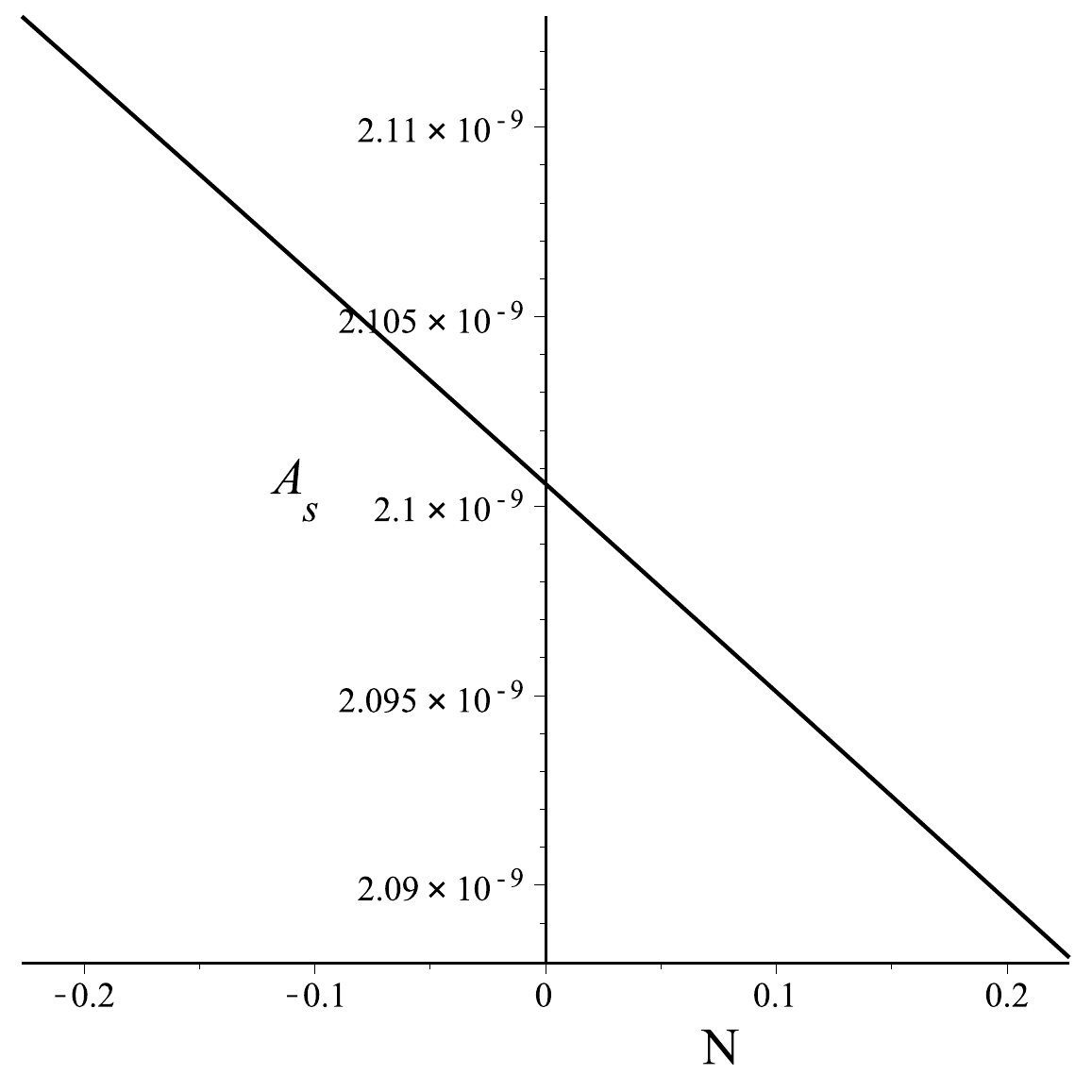}
\caption{The values of the inflationary parameters $n_s$ (left), $r$ (center), and $A_s$ (right). The model parameters are given in (\ref{modelparam}).}
\label{inlparamN}
\end{figure}

Results of numerical integration of the evolution equations~(\ref{DYNSYSTEMN}) with the following values of parameters:
\begin{equation}
\label{modelparam}
   \delta=2.5\cdot 10^{-4},\quad U_0=0.8,\quad  \chi_0=2\,,\quad C=0.00044,\quad  d=0.0005,\quad m=2.3084\cdot10^{-5}\, M_\mathrm{Pl},
\end{equation}
are presented in Figs.~\ref{Trajectory}--\ref{inlparamN}.
In Fig.~\ref{Trajectory}, we demonstrate a typical evolution of the Hubble parameters and scalar field in the model proposed.
In Fig.~\ref{slrparamN}, one can see two slow-roll stages and the violation of the slow-roll regime between them. The values of the inflationary parameters $n_s$, $r$, and $A_s$ at $N=0$ are presented in Fig.~\ref{inlparamN}.

The choice of model parameters in formula (\ref{modelparam}) is not unique. Tables \ref{TablParam1} and \ref{TablParam2} show that different values of these parameters can lead to viable inflationary scenarios with different inflationary parameters. The duration of the first stage of inflation, $N_*$, and the total duration of inflation, $N_{e}$, can vary depending on the chosen values of the parameters.
The field $\phi_0$ is determined by the condition $n_s(\phi_0) = 0.974$. After that, the parameter $m$ is selected such that $A_s(\phi_0) = 2.1 \cdot 10^{-9}$.

 The value of the parameter $\delta$ is important for both stages of inflation  as shown in Table~\ref{TablParam1}, whereas the value of parameter $d$ determines duration of the second stage of inflation only. So, choosing $\delta$ and fixing other parameters, one can get suitable inflationary parameters, after this we can fix $\delta$ and change $d$ to get scenarios with the same inflationary parameters but different duration of the second stage and, therefore, different values of the PBH mass (see Table~\ref{TablParam2}). For example, at $\delta=2.0\cdot10^{-4}$ and $d=0.001$, the value of $N_{e}$ is too large, $N_{e}=70.3$. To decrease this value we increase value of $d$. At $d=0.005$ and values of other parameters as in Table~\ref{TablParam1}, we obtain $N_{e}=63.3$ and $M_{PBH}\approx1.63\cdot10^{-15}M_{\bigodot}$. The values of the inflationary parameters $n_s$ and $r$ as well as the value of $N_*$ remain the same as in the first row of  Table~\ref{TablParam1}.
It is easy to see that the model does not contradict the observational data for the chosen values of parameter.

\begin{table}
  \centering
\begin{tabular}{|c|c|c|c|c|c|c|c|}
  \hline
  $\delta$ & ${m}/{M_\mathrm{Pl}}$ & ${\phi_0}/{M_\mathrm{Pl}}$ & $n_s$ & $r$ & $N_{*}$& $N_{e}$ & $M_{PBH}/M_{\bigodot}$\\
  \hline
  $2.0\cdot10^{-4}$ & $2.101\cdot10^{-5}$ & $5.088$ & $0.974$ & $0.0105$ & $42.2$ & $70.3$ & $2.14\cdot10^{-9}$  \\
  $2.1\cdot10^{-4}$ & $2.144\cdot10^{-5}$ & $5.071$ & $0.974$ & $0.0109$ & $41.5$ & $66.6$ & $4.25\cdot10^{-12}$ \\
  $2.3\cdot10^{-4}$ & $2.235\cdot10^{-5}$ & $5.027$ & $0.974$ & $0.0118$ & $39.9$ & $61.2$ & $1.53\cdot10^{-15}$ \\
  $2.5\cdot10^{-4}$ & $2.308\cdot10^{-5}$ & $5.007$ & $0.974$ & $0.0126$ & $39.1$ & $58.1$ & $2.11\cdot10^{-17}$ \\
  $2.7\cdot10^{-4}$ & $2.384\cdot10^{-5}$ & $4.981$ & $0.974$ & $0.0134$ & $38.1$ & $55.6$ & $8.28\cdot10^{-19}$\\
  $2.9\cdot10^{-4}$ & $2.4595\cdot10^{-5}$ & $4.952$ & $0.974$ & $0.0143$ & $37.1$ & $53.3$& $5.82\cdot10^{-20}$ \\
  $3.1\cdot10^{-4}$ & $2.532\cdot10^{-5}$ & $4.927$ & $0.974$ & $0.0152$ &  $36.2$ & $51.6$ & $9.33\cdot10^{-21}$ \\
 \hline
\end{tabular}
  \caption{The dependence of the inflation parameter $r$, the duration of the first stage of inflation $N_{*}$, the total duration of inflation $N_{e}$, and the PBH mass on the model parameter $\delta$.
  The value of the parameter $m$ is fixed by the condition $A_s(\phi_0)=2.1\cdot 10^{-9}$. Other model parameters are chosen as follows:  $U_0=0.8\,$,  $\chi_0=2\,$, $C=0.00044\,$, $d=0.001\,$.
  \label{TablParam1}}
\end{table}

\begin{table}
  \centering
\begin{tabular}{|c|c|c|c|c|c|c|}
  \hline
  $d$ & $N_{e}$ &$N_{e}-N_{*}$ & $M_{PBH}/M_\mathrm{Pl}$ & $M_{PBH}/M_{\bigodot}$ & $M_{PBH}/g$ &$H_{e}/{M_\mathrm{Pl}}$ \\
  \hline
  $0.0012$& $57.7$ & $18.6$ & $3.54\cdot10^{21}$ & $7.72\cdot10^{-18}$ & $1.56\cdot10^{16}$& $4.04\cdot 10^{-6}$\\
  $0.0010$& $58.1$ & $19.1$ & $9.70\cdot10^{21}$ & $2.11\cdot10^{-17}$ & $4.27\cdot10^{16}$& $4.01\cdot 10^{-6}$\\
  $0.0008$ &$58.8$ & $19.7$ & $3.21\cdot10^{22}$ & $7.00\cdot10^{-17}$ & $1.41\cdot10^{17}$ & $4.02\cdot 10^{-6}$ \\
  $0.0005$ & $60.0$ & $20.9$ & $3.53\cdot 10^{23}$ & $7.70\cdot 10^{-16}$ & $1.55\cdot 10^{18}$  & $4.02\cdot 10^{-6}$ \\
  $0.0003$ & $61.4$ & $22.3$ & $5.85\cdot10^{24}$ & $1.28\cdot10^{-14}$ & $2.57\cdot10^{19}$ & $4.00\cdot 10^{-6}$ \\
  $0.0002$ & $62.5$ & $23.5$ & $6.42\cdot 10^{25}$& $1.40\cdot10^{-13}$& $2.82\cdot 10^{20}$ & $4.02\cdot  10^{-6}$\\
  $0.00015$& $63.3$ & $24.2$ & $2.62\cdot 10^{26}$& $5.71\cdot10^{-13}$& $1.15\cdot 10^{21}$ & $4.01\cdot 10^{-6}$\\
  $0.0001$ & $64.4$ & $25.3$ & $2.36\cdot 10^{27}$ & $5.14\cdot 10^{-12}$ & $1.04\cdot10^{22}$ & $4.00\cdot 10^{-6}$  \\
  $0.00008$ & $64.9$ & $25.9$ & $7.77\cdot 10^{27}$ & $1.69\cdot 10^{-11}$ & $3.42\cdot10^{22}$ & $4.04\cdot 10^{-6}$  \\
  \hline
\end{tabular}
\caption{The dependence of the duration of inflation $N_{e}$ and the PBH mass $M_{PBH}$ on the model parameter~$d$. Inflationary parameters, $n_s=0.974$ and $r=0.0126$, as well as the duration of the first stage of inflation $N_{*}\approx 39$ are independent of $d$. Other model parameters are chosen as follows:  $U_0=0.8\,$, $\delta=2.5\cdot 10^{-4}$,\, $\chi_0=2\,$, $C=0.00044\, $, $m=2.3084\cdot10^{-5}\, M_\mathrm{Pl}$.
  \label{TablParam2}}
\end{table}

If PBH mass belongs to the following interval $10^{-17}\,M_{\bigodot}\leqslant M_{PBH}\leqslant 10^{-12} M_{\bigodot}$, where $M_{\bigodot}$ is the Solar mass, then this PBH can be considered as a part of dark matter~\cite{Ozsoy:2023ryl}. As shown in Table~\ref{TablParam2}, the proposed $F(R,\chi)$ model gives the masses of the PBH from this interval at $0.0001\leqslant d\leqslant 0.0010$.

\section{Conclusions}

In this paper, we propose the $F(R,\chi)$ gravity models,  which unify inflation and PBH formation. Using the conformal transformation of the metric, we get the corresponding chiral cosmological model with two scalar fields. We have found such values of the model parameters, at which the model constructed is in a good agreement with the ACT observation data and is suitable for describing the formation of PBHs. The choice of the model parameters allows us to obtain such black hole masses that the resulting PBHs could be considered as dark matter candidates.

We should note that the choice of the potential $V_E(\phi,\chi)$ is not determined by  any specific particle physics model, so this model can be considered as a toy model. We hope that the proposed model will be useful in constructing more realistic models that unify inflation and PBH production and are based on particle physics. For further investigations, it would be interesting to consider processes during and after inflation in the Jordan frame, generalizing the methods of slow-roll approximation construction proposed in Refs.~\cite{Pozdeeva:2025ied,Ketov:2025cqg} to models with a few scalar fields.

\section*{FUNDING}
This study was conducted within the scientific program of the National Center for Physics and Mathematics, section 5 'Particle Physics and Cosmology'. Stage 2023--2025.

\section*{CONFLICT OF INTEREST}

The authors of this work declare that they have no conflicts of interest.


\bibliography{PV_FR_CiteAll}{}
\bibliographystyle{utphys}

\end{document}